\documentclass[aps,pra,preprint,showpacs,amsmath,superscriptaddress]{revtex4-1}

\usepackage{hyperref}
\usepackage{siunitx}

\newcommand{\tm}{\mathrm{TM}}
\newcommand{\te}{\mathrm{TE}}
\newcommand{\F}{\mathcal{F}}
\newcommand{\kb}{k_{B}}
\newcommand{\Li}[2]{\mathrm{Li}_{#1}\left(#2\right)}
\newcommand{\Dt}{\Delta_{T}}
\renewcommand{\coth}[2][]{
\,\mathrm{coth}^{#1}\left(#2\right)\,
}
\renewcommand{\sinh}[2][]{
\,\mathrm{sinh}^{#1}\left(#2\right)\,
}
\renewcommand{\cosh}[2][]{
\,\mathrm{cosh}^{#1}\left(#2\right)\,
}
\renewcommand{\cos}[2][]{
\,\mathrm{cos}^{#1}\left(#2\right)\,
}
\renewcommand{\sin}[2][]{
\,\mathrm{sin}^{#1}\left(#2\right)\,
}

\begin{document}


\title{Analytic results for the Casimir free energy between ferromagnetic metals}


\author{G.~L. Klimchitskaya}
\affiliation{
    Central Astronomical Observatory at Pulkovo of the Russian Academy of
    Sciences,
    St. Petersburg,
    196140,
    Russia
}
\affiliation{
    Institute of Physics, Nanotechnology and Telecommunications,
    St. Petersburg State Polytechnical University,
    St. Petersburg,
    195251,
    Russia
}

\author{C.~C. Korikov}
\affiliation{
    Institute of Physics, Nanotechnology and Telecommunications,
    St. Petersburg State Polytechnical University,
    St. Petersburg,
    195251,
    Russia
}


\begin{abstract}
We derive perturbation analytic expressions for the Casimir free energy
and entropy between two dissimilar ferromagnetic plates which are applicable at
arbitrarily low temperature. The dielectric properties of metals are described
using either the nondissipative plasma model or the Drude model taking into
account the dissipation of free charge carriers. Both cases of constant and
frequency-dependent magnetic permeability are considered. It is shown that for
ferromagnetic metals described by the plasma model the Casimir  entropy goes
to zero when the temperature vanishes, i.e., the Nernst heat theorem is satisfied.
For ferromagnetic metals with perfect crystal lattices described by the Drude
model the Casimir entropy goes to a nonzero constant depending on the parameters
of a system with vanishing temperature, i.e., the Nernst heat theorem is violated.
This constant can be positive which is quite different from the earlier
investigated  case of two nonmagnetic metals.
\end{abstract}


\maketitle

\section{\label{sec:intro}Introduction}

In the last few years physical phenomena caused by quantum fluctuations of
the electromagnetic field attracted much attention in both fundamental physics
and technological applications. Among other fluctuation phenomena, the van der
Waals~\cite{c01} and Casimir~\cite{c02} forces occupy a highly important place because they
manifest itself as a macroscopic interaction between closely spaced material
bodies. The unified theory of the van der Waals and Casimir forces based on
quantum electrodynamics was developed by Lifshitz~\cite{c03} sixty years ago, but sufficiently
precise measurements have been performed only recently (see Refs.~\cite{c04,c05,c06} for a review).
These measurements opened prospective applications of the Casimir forces in
nanotechnology~\cite{c07,c08}, Bose-Einstein condensation~\cite{c09},
semiconductors~\cite{c10,c11,c12,c13,c14,c15,c16}, phase transitions~\cite{c17},
graphene microstructures~\cite{c18} etc.

Although the Lifshitz theory turned out to be very useful for interpretation
of the measurement data, the most precise experiments~\cite{c19,c20,c21,c22,c23} using the
configuration of nonmagnetic metallic test bodies demonstrated a disagreement with
theoretical predictions if the relaxation properties of conduction electrons are
taken into account in calculations. The same measurement data were found in
agreement with theoretical predictions of the Lifshitz theory if the relaxation
properties of conduction electrons are disregarded~\cite{c19,c20,c21,c22,c23}. Simultaneously,
it was shown~\cite{c24,c25,c26} that for metals with perfect crystal lattices the Casimir
entropy found in the framework of the Lifshitz theory violates the third law of
thermodynamics (the Nernst heat theorem) if the relaxation properties of conduction
electrons are taken into account. With omitted contribution of relaxation properties,
the Nernst heat theorem is satisfied. The relaxation properties of free electrons
are well described by the Drude model. The theoretical approach taking these
properties into account in calculations of the Casimir force is known as the Drude
model approach. The plasma of conduction electrons with no dissipation is
described by the plasma model. Usually it is applicable at high frequencies,
which are much larger than the relaxation frequency. The theoretical approach
disregarding the relaxation properties of conduction electrons in calculations
of the Casimir force is called the plasma model approach.

A disagreement of the measurement data with the Drude model approach and the
violation of the Nernst heat theorem in this approach are puzzling and created
a discussion in the literature. Specifically, it was underlined~\cite{c27,c28} that for
real metals with some fraction of impurities the Casimir entropy jumps to zero at
sufficiently low temperature, i.e., the Nernst heat theorem is formally restored.
This, however, does not provide a satisfactory explanation for a puzzle~\cite{c29}.
It was also shown~\cite{c30} that at large separations, where the Casimir force is classical,
the plasma model approach violates the Bohr-van Leeuwen theorem, whereas the
Drude model approach is consistent with it. Taking into account that at short
separations below a micrometer, where measurements are most precise, the relative
difference in theoretical predictions of the Drude and plasma model approaches
is equal to only a few percent, a more definitive experimental evidence is highly
desirable.

Such an evidence was provided by recent experiments on measuring the gradient
of the Casimir force between ferromagnetic metals~\cite{c31,c32,c33}. Here, the relative
difference between the predictions of two approaches is either almost zero (Au-Ni system)
or has an opposite sign with respect to the Au-Au system. These permit to exclude
the role of any possible systematic effect that could plague the theory-experiment
comparison. In Refs.~\cite{c31,c32,c33} the plasma model approach was again confirmed.
Further value of ferromagnetic (or, synonymously, magnetic) metals for
the Casimir physics is that they provide the possibility to perform an
experiment where theoretical predictions of the Drude and plasma model
approaches differ not by a few percent but by a factor of 1000~\cite{c34}
(see also Ref.~\cite{c34a}). The first data sets
of such an experiment are already reported. They indicate conclusively that
the Drude model approach is excluded, whereas the plasma model approach is
in agreement with the data~\cite{c35}.

Taking into account the crucial importance of magnetic properties for the
resolution of a puzzle formulated above, in this paper we derive the analytic
expressions for the Casimir free energy and entropy in the configuration of
two parallel plates at temperature $T$ made of dissimilar magnetic metals.
All the results below are obtained in the framework of the Drude and plasma model
approaches in the form of perturbation expansions. In the plasma model approach,
the used small parameters are the relative penetration depths of the electromagnetic
oscillations into metals multiplied by the square roots of static magnetic
permeabilities. In the case of the Drude model approach the ratios of the relaxation
frequencies to the first Matsubara frequency serve as additional small parameters.
The derivations are first performed assuming that the magnetic permeabilities are
static and then generalized for the case of frequency-dependent permeability.

The obtained expressions are used to investigate the low-temperature behavior
of the Casimir entropy for magnetic metals. It is shown that for the plasma
model the Casimir entropy goes to zero when the temperature vanishes, i.e.,
the Nernst heat theorem is satisfied. For magnetic metals with perfect crystal
lattices described by the Drude model, the Casimir entropy
goes to a nonzero limit depending on the parameters of a system when
the temperature goes to zero, i.e., the Nernst heat theorem is violated.
We prove that at zero temperature the Casimir entropy of magnetic metals
described by the Drude model can be positive. This is different from
the case of nonmagnetic Drude metals where the Casimir energy at zero
temperature is always negative.

The paper is organized as follows. In Sec.~\ref{sec:plasma_model} we derive
perturbation expansions for the Casimir free energy and entropy calculated
using the plasma model approach in the case on constant magnetic permeabilities
and investigate the limit of zero temperature. In Sec.~\ref{sec:drude_model}
the same is done when the Drude model approach is used in calculations.
Section~\ref{sec:freq_dependence} contains generalization of the obtained
results for the case of frequency-dependent magnetic permeabilities.
In Sec.~\ref{sec:conc} the reader will find out conclusions and discussion.

\section{\label{sec:plasma_model}Perturbation expansions of the Casimir free
energy and entropy in the plasma model}

We consider the configuration of two $(n=1,2)$ parallel thick plates
(semispaces) at a separation $a$ made of dissimilar magnetic metals
characterized by the frequency-dependent dielectric permittivities $\epsilon^{(n)}(\omega)$
and magnetic permeabilities $\mu^{(n)}(\omega)$. The Lifshitz formula for
the Casimir free energy per unit area of plates written in terms of dimensionless
variables takes the form~\cite{c02,c03}
\begin{widetext}
\begin{equation}
\F(a,T) = \frac{\kb T}{8 \pi a^2} \sum_{l=0}^{\infty}\phantom{}^{'}
    \int_{\zeta_l}^{\infty} y\,d y
    \sum_{\alpha}
        \ln
            \left[
                1-r_{\alpha}^{(1)}(i\zeta_l,y)r_{\alpha}^{(2)}(i\zeta_l,y) e^{-y}
            \right].
\label{eq:01}
\end{equation}
\end{widetext}
Here, $\kb$ is the Boltzmann constant, $T$ is the temperature and $\zeta_l$
are the dimensionless Matsubara frequencies connected with the dimensional ones
$\xi_l=2\pi l  \kb T / \hbar$ by the relation $\zeta_l = \xi_l/\omega_c$ where $\omega_c=c/(2a)$.
The prime in the first sum on the right-hand side of Eq.~\eqref{eq:01} means that the term with $l=0$ is divided by two.
The second sum is over two independent polarizations of the electromagnetic field,
transverse magnetic $(\alpha=\text{TM})$ and transverse electric $(\alpha=\text{TE})$.
The reflection coefficients in Eq.~\eqref{eq:01} calculated at the imaginary Matsubara
frequencies are given by~\cite{c02}
\begin{align}
\begin{split}
r_{\tm}(i \zeta_l,y) &=
    \frac{
        \epsilon_l^{(n)}y-\sqrt{y^2+[\epsilon_l^{(n)}\mu_l^{(n)}-1]\zeta_l^2}
    }
    {
        \epsilon_l^{(n)}y+\sqrt{y^2+[\epsilon_l^{(n)}\mu_l^{(n)}-1]\zeta_l^2}
    }, \\
r_{\te}(i \zeta_l,y) &=
    \frac{
        \mu_l^{(n)}y-\sqrt{y^2+[\epsilon_l^{(n)}\mu_l^{(n)}-1]\zeta_l^2}
    }
    {
        \mu_l^{(n)}y+\sqrt{y^2+[\epsilon_l^{(n)}\mu_l^{(n)}-1]\zeta_l^2}
    },
\end{split}
\label{eq:02}
\end{align}
where $\epsilon_l^{(n)}\equiv\epsilon^{(n)}(i \zeta_l \omega_{c})$ and $\mu_l^{(n)}\equiv\mu^{(n)}(i \zeta_l \omega_{c})$.

In this section we consider the dielectric permittivity of the plasma model
which describes the nondissipative gas of free electrons~\cite{c36}. At the imaginary
Matsubara frequencies the dielectric permittivities of both plates in the
framework of the plasma model are
\begin{equation}
\epsilon_l^{(n)} = 1 + \left(\frac{{\omega_{p}^{(n)}}}{\xi_l}\right)^2
=
1+\left(\frac{\widetilde{\omega}_{p}^{(n)}}{\zeta_l}\right)^2,
\label{eq:03}
\end{equation}
where $\omega_{p}^{(n)}$ are the plasma frequencies for the metals of the plates
and $\widetilde{\omega}_{p}^{(n)} \equiv \omega_{p}^{(n)}/\omega_{c}$.

Now we use the calculation procedure developed in Ref.~\cite{c37} in the case of
nonmagnetic metals. In this and in the next section we assume constant magnetic
permeabilities $\mu_l^{(n)}=\mu_0^{(n)}$. Using the Poisson summation formula adapted
for the case of even functions~\cite{c02}, Eq.~\eqref{eq:01} can be rewritten in the form

\begin{equation}
\F(a,T) = \frac{\hbar c}{16 \pi^2 a^3} \sum_{l=0}^{\infty}\phantom{}^{'}
\int_{0}^{\infty} y\,d y
\int_{0}^{y}
\,d \zeta \cos{lt\zeta}
F(\zeta,y),
\label{eq:04}
\end{equation}
where $t=T_{\text{eff}}/T \equiv \hbar c/(2a \kb T)$ and

\begin{align}
F(\zeta,y) = \sum_{\alpha}
    \ln
        \left[
            1-r_{\alpha}^{(1)}(i \zeta, y)r_{\alpha}^{(2)}(i \zeta, y) e^{-y}
        \right].
\label{eq:05}
\end{align}

The term of Eq.~\eqref{eq:05} with $l=0$ describes the Casimir energy per unit area $E(a)$ at zero temperature. The terms with $l\ge 1$ represent the thermal correction to it. Then Eq.~\eqref{eq:04} can be written as
\begin{equation}
\F(a,T) = E(a) + \Dt \F(a,T),
\label{eq:06}
\end{equation}
where
\begin{equation}
\Dt \F(a,T) = \frac{\hbar c}{16 \pi^2 a^3} \sum_{l=1}^{\infty}
    \int_{0}^{\infty} y \, d y
    \int_{0}^{y} \, d \zeta \cos{l t \zeta} F(\zeta,y).
\label{eq:07}
\end{equation}

Now we consider separation distances between the plates satisfying a condition

\begin{equation}
a \gg \lambda_{p}^{(n)} = \frac{2 \pi c}{\omega_{p}^{(n)}},
\label{eq:08}
\end{equation}
where $\lambda_{p}^{(n)}$ are the plasma wavelengths for both plates. In this
separation region the characteristic frequency $\omega_{c}$ is much smaller
than $\omega_{p}^{(n)}$. We also assume that  separation distances are so large
that the following equality is satisfied.

\begin{equation}
\Lambda \equiv
\frac{\lambda_{p}^{(1)} \sqrt{\mu_0^{(1)}}+\lambda_{p}^{(2)} \sqrt{\mu_0^{(2)}}}
{4 \pi a}
\ll 1.
\label{eq:09}
\end{equation}
For example, for Ni used in experiments on measuring the Casimir force between
magnetic metals~\cite{c31,c32,c33} $\mu_0\approx 110$ and $\lambda_p/(2\pi) \approx 40$~nm.
Thus, for two Ni plates the
inequality~\eqref{eq:09} is satisfied for $a>2\,\mu$m. Expanding Eq.~\eqref{eq:05} in
powers of a small parameter~\eqref{eq:09}, we obtain

\begin{equation}
F(\zeta,y) = 2 \ln\left(1-e^{-y}\right)
+
2 \frac{\zeta^2+y^2}{y (e^{y}-1)}\Lambda
-\frac{2e^{-y}}{\left(1-e^{-y}\right)^2} \frac{\zeta^4+y^4}{y^2} \Lambda^2.
\label{eq:10}
\end{equation}

This equation can be substituted in the right-hand side of Eqs.~\eqref{eq:04} and~\eqref{eq:07}
and all integrals with respect to $\zeta$ can be calculated explicitly with the result
\begin{equation}
\int_{0}^{y} \, d \zeta \cos{lt\zeta} F(\zeta,y)
=
A_l^{(0)}(y) + A_l^{(1)}(y)\Lambda + A_l^{(2)}(y) \Lambda^2,
\label{eq:11}
\end{equation}
where the functions $A_l^{(0)}$, $A_l^{(1)}$ and $A_l^{(2)}$ are given by
\begin{align}
\begin{split}
A_l^{(0)}(y) = & \frac{2}{lt} \ln\left(1-e^{-y}\right)\sin{lty},\\
A_l^{(1)}(y) = &-\frac{4}{y \left(e^{y}-1\right)}
\left[
    \frac{\sin{lty}}{l^3 t^3}-\frac{y\cos{lty}}{l^2 t^2} - \frac{y^2\sin{lty}}{lt}
\right],\\
A_l^{(2)}(y) = &-\frac{4 e^{-y}}{y^2 \left(1-e^{-y}\right)^2}
\left[
    \frac{12 \sin{lty}}{l^5 t^5} - \frac{12 y \cos{lty}}{l^4 t^4}
    -
    \right. \\
    -& \left.
    \frac{6 y^2 \sin{lty}}{l^3 t^3} + \frac{2 y^3 \cos{lty}}{l^2 t^2} + \frac{y^4 \sin{lty}}{l t}
\right].
\label{eq:12}
\end{split}
\end{align}
After substitution of Eqs.~\eqref{eq:11} and~\eqref{eq:12} in Eqs.~\eqref{eq:04} and~\eqref{eq:07},
the integrals with respect to $y$  are also calculated in the form
\begin{eqnarray}
&&
\int_{0}^{\infty} y \,dy \int_{0}^{y}
\left[
    A_l^{(0)}(y) + A_l^{(1)}(y)\Lambda + A_l^{(2)}(y) \Lambda^2
\right]
\nonumber\\
&&~~~~=
B_l^{(0)}(t) + B_l^{(1)}(t)\Lambda + B_l^{(2)}(t) \Lambda^2,
\label{eq:13}
\end{eqnarray}
where for the functions $B_l^{(0)}$, $B_l^{(1)}$ and $B_l^{(2)}$ we find
\begin{align}
\begin{split}
B_l^{(0)}(t) &= 2
    \left[\frac{1}{l^4t^4}-\frac{\pi \coth{\pi lt}}{2 l^3t^3}-\frac{\pi^2}{2 l^2 t^2 \sinh[2]{\pi lt}}
\right],\\
B_l^{(1)}(t) &= -4
\left[
    \frac{\pi}{l^3t^3} \coth{\pi lt} - \frac{4}{l^4 t^4} + \frac{\pi^2}{l^2 t^2}
    \frac{1}{\sinh[2]{\pi lt}}
    \right. \\ & +
    \left.
    \frac{2\pi^3}{l t} \frac{\cosh{\pi lt}}{\sinh[2]{\pi lt}}
\right],\\
B_l^{(2)}(t) &= 2
\left\{
\frac{\pi}{l^5t^5}
+
\frac{2 \pi^4}{\sinh[2]{\pi lt}}
\left[
    \frac{3\coth{\pi lt}\sinh[2]{\pi lt}}{\pi^3 l^3t^3}
    \right. \right. \\ & -
    \left.
    2 \coth[2]{\pi lt }
    -
    \frac{1}{\sinh[2]{\pi lt}}
    +
    \frac{\coth{\pi lt}}{\pi lt} - \frac{1}{\pi^2 l^2t^2}
\right]
\\ & +
\frac{12}{l^4t^4}
\ln\left(1-e^{-2 \pi lt}\right)
-
\frac{6 \pi}{l^3t^3}
-
\frac{24 \pi}{l^3 t^3} \frac{1}{e^{2\pi lt}-1}
\\ & -
\left.
\frac{6}{\pi l^5 t^5}
\Li{2}{e^{-2\pi lt}}
\right\}.
\label{eq:14}
\end{split}
\end{align}
Here, $\Li{n}{z}$ is the polylogarithm function.

As a result, the thermal correction~\eqref{eq:07} takes the following explicit form:
\begin{equation}
\Dt \F(a,T) = \frac{\hbar c}{16 \pi^2 a^3} \sum_{l=1}^{\infty}
\left[
B_l^{(0)}(t) + B_l^{(1)}(t)\Lambda + B_l^{(2)}(t) \Lambda^2
\right],
\label{eq:15}
\end{equation}
where the temperature-dependent coefficients $B_l^{(0)}$, $B_l^{(1)}$ and
$B_l^{(2)}$ are given in Eq.~\eqref{eq:14}. Calculating the negative derivative
of Eq.~\eqref{eq:15} with respect to temperature, one obtains the explicit expression for the
Casimir entropy.

For our purposes it is desirable to find the asymptotic behavior of the thermal
correction~\eqref{eq:15} at arbitrarily low temperatures $T \ll T_{\text{eff}}$. This corresponds
to the condition $t \gg 1$. Keeping only the largest of the exponentially small
contributions in Eq.~\eqref{eq:15}, taking into account that
\begin{equation}
\lim_{|z|\to 0} \Li{n}{z} = z
\label{eq:16}
\end{equation}
and performing all necessary summations, one obtains
\begin{align}
\begin{split}
\Dt \F(a,T) & = -\frac{\hbar c}{8 \pi a^3}
\left\{
    \frac{\zeta_{R}(3)}{2 t^3}
    -
    \frac{\pi^3}{90 t^4}
    +
    \frac{2 \pi}{t^2}
    e^{-2\pi t}
    \right.\\ &  +
    \Lambda
    \left[
        \frac{\zeta_{R}(3)}{t^3}
        -
        \frac{2\pi^3}{45 t^4}
        +\frac{8\pi^2}{t}
        e^{-2\pi t}
    \right]
    \\ &  \left. -
    \Lambda^2
    \left[
        \frac{\zeta_{R}(5)}{t^5}
        -
        16 \pi^3
        e^{-2\pi t}
    \right]
\right\},
\label{eq:17}
\end{split}
\end{align}
where $\zeta_{R}(z)$ is Riemann zeta function.

Equation~\eqref{eq:17} gives the possibility to find the asymptotic behavior of
the Casimir entropy
\begin{equation}
S(a,T) = -\frac{\partial \Dt \F(a,T)}{\partial T}
\label{eq:18}
\end{equation}
when $T$ goes to zero. Omitting the exponentially small contributions, from
Eqs.~\eqref{eq:17} and~\eqref{eq:18} we arrive at
\begin{align}
\begin{split}
S(a,T) & = \frac{\kb \tau^2}{16 a^2 \pi^3}
\left\{
    \frac{3\zeta_{R}(3)}{2}
    -
    \frac{\pi^2}{45}\tau
    \right. \\ & \left. +
    \Lambda
    \left[
        3\zeta_{R}(3) - \frac{4 \pi^2}{15} \tau
    \right]
    -
    \Lambda^2
    \frac{5 \zeta_{R}(5)}{4 \pi^2}\tau^2
\right\},
\label{eq:19}
\end{split}
\end{align}
where we have introduced the dimensionless temperature
\begin{equation}
\tau = 2\pi \frac{T}{T_{\text{eff}}} = \frac{2 \pi }{t} = \frac{4 \pi a \kb T}{\hbar c}.
\label{eq:20}
\end{equation}

As is seen from Eq.~\eqref{eq:19}, the Casimir entropy goes to zero when the
temperature vanishes in accordance with the Nernst heat theorem. One can conclude
that the Lifshitz theory combined with the plasma model provides thermodynamically
consistent description of the Casimir interaction between magnetic metals
(previously this statement was proved for the case of nonmagnetic metal
plates~\cite{c24,c25,c26}).

\section{\label{sec:drude_model}Perturbation expansions of the Casimir free energy and entropy in the Drude model}

Here, we consider the Casimir free energy~\eqref{eq:01} with reflection
coefficients~\eqref{eq:02}, as given by the Lifshitz theory~\cite{c02,c03}.
However, instead of the dielectric permittivity of the plasma model~\eqref{eq:03},
we use the dielectric permittivity of the Drude model at the imaginary Matsubara frequencies
\begin{equation}
\epsilon_l^{(n)} = 1 + \frac{\left({\omega_{p}^{(n)}}\right)^2}{\xi_l \left[
\xi_l + \gamma^{(n)}(T)
\right]} = 1+\frac{\left(\widetilde{\omega}_{p}^{(n)}\right)^2}{\zeta_l\left[
\zeta_l
+
\widetilde\gamma^{(n)}(T)
\right]}.
\label{eq:21}
\end{equation}
In this equation, $\gamma^{(n)}(T)$ are the relaxation parameters (relaxation frequencies)
of the metals of plates. The relaxation parameters depend on the temperature and
for perfect crystal lattices go to zero faster than the first power of $T$ with
vanishing temperature~\cite{c26,c38}. The dimensionless relaxation parameter
is defined as $\widetilde\gamma^{(n)}(T) = \gamma^{(n)}(T)/\omega_c$.
As is seen from Eq.~\eqref{eq:21}, at any $\zeta_l \neq 0$
the plasma model~\eqref{eq:03} can be considered as a limiting case of
the Drude model~\eqref{eq:21} when $\gamma^{(n)}$ goes to zero. Generally this statement is,
however, incorrect because in the limiting case $\gamma^{(n)} \to 0$
the Drude model along the real frequency axis possesses a singularity
proportional to $\delta(\omega)$~\cite{c39}.

Now we substitute the dielectric permittivity~\eqref{eq:21} in the reflection
coefficients~\eqref{eq:02}. For convenience in calculations, below we supply all
quantities found in Sec.~\ref{sec:plasma_model} using the plasma model~\eqref{eq:03}
with an index $p$, and the respective quantities found using the Drude model~\eqref{eq:21}
with an index $D$. Specifically, for the reflection coefficients calculated
at zero Matsubara frequency using the two models, one obtains
\begin{align}
\begin{split}
&r_{\tm}^{D(n)} (0,y) = r_{\tm}^{p(n)} (0,y) = 1,\,
r_{\te}^{D(n)} (0,y) = \frac{\mu_0^{(n)}-1}{\mu_0^{(n)}+1} \equiv r_{\mu}^{(n)},\\
&
r_{\te}^{p(n)} (0,y) =
\frac{\mu_0^{(n)}y-\sqrt{(\widetilde\omega_p^{(n)})^2 \mu_0^{(n)} + y^2}}
{\mu_0^{(n)}y+\sqrt{(\widetilde\omega_p^{(n)})^2 \mu_0^{(n)} + y^2}}.
\label{eq:22}
\end{split}
\end{align}

For the calculation of the Casimir free energy $\F_{D}(a,T)$ using the Drude model,
it is useful to present it identically as
\begin{equation}
\F_{D}(a,T)=\F_{p}(a,T)+\F_{D}(a,T)-\F_{p}(a,T)
\label{eq:23}
\end{equation}
and to separate the zero-frequency terms of the last two quantities in the following way:
\begin{align}
\begin{split}
    \F_{D}(a,T)&=\F_{p}(a,T)
    +
    \frac{\kb T}{16 \pi a^2}
    \\ & \times
    \int_{0}^{\infty} y \, dy
    \left\{
        \ln\left[1-r^{D(1)}_{\te}(0,y)r^{D(2)}_{\te}(0,y)e^{-y}\right]
        \right. \\ & - \left.
        \ln\left[1-r^{p(1)}_{\te}(0,y)r^{p(2)}_{\te}(0,y)e^{-y}\right]
    \right\}
    +
    \frac{\kb T}{8 \pi a^2}
    \\ & \times
    \sum_{l=1}^{\infty}
    \int_{\zeta_l}^{\infty} y \, dy
    \sum_{\alpha}
    \left\{
    \ln\left[1-r^{D(1)}_{\alpha}(i \zeta_l,y)r^{D(2)}_{\alpha}(i \zeta_l,y)e^{-y}\right]
        \right. \\ & - \left.
        \ln\left[1-r^{p(1)}_{\alpha}(i \zeta_l,y)r^{p(2)}_{\alpha}(i \zeta_l,y)e^{-y}\right]
    \right\}.
\label{eq:24}
\end{split}
\end{align}
Note that due to Eq.~\eqref{eq:22} the TM contributions at zero Matsubara
frequency in both models cancel each other.

We expand the products of the Drude reflection coefficients for two plates up
to the first powers in small parameters $\widetilde\gamma^{(n)}(T)/\zeta_l$. Introducing also the notation
$\beta_n \equiv \lambda_p^{(n)}/(4\pi a)\ll 1$, for the TM polarization we obtain
\begin{eqnarray}
&&
r^{D(1)}_{\tm} (i \zeta_l,y) r^{D(2)}_{\tm}(i \zeta_l,y)
=
r^{p(1)}_{\tm} (i \zeta_l,y) r^{p(2)}_{\tm}(i \zeta_l,y)
\nonumber\\
&&~~~-
\frac{\widetilde{\gamma}^{(1)}(T)}{\zeta_l}
R^{(1)}_{\tm} (i \zeta_l,y)
-
\frac{\widetilde{\gamma}^{(2)}(T)}{\zeta_l}
R^{(2)}_{\tm} (i \zeta_l,y),
\label{eq:25}
\end{eqnarray}
where  the expansion coefficients are given by
\begin{eqnarray}
&&
R_{\tm}^{(n)} (i \zeta_l,y) =
\frac{
    \beta_n\zeta_l^2 y\left\{\mu_0^{(n)}+\beta_n^2\left[
    \zeta_l^2\mu_0^{(n)}+2\left(y^2-\zeta_l^2\right)
    \right]\right\}
}
{
    \sqrt{\beta_n^2 y^2 + \beta_n^2 \zeta_l^2(\mu_0^{(n)}-1)+\mu_0^{(n)}}
}
\nonumber\\
&&~~~ \times
\frac{
r_{\tm}^{p(1)}(i \zeta_l,y)r_{\tm}^{p(2)}(i \zeta_l,y)
}
{
\beta_n^2 \zeta_l^2 \left\{2 y^2 - \zeta_l^2 \left[\beta_n^2\zeta_l^2(\mu_0^{(n)}-1)+\mu_0^{(n)}\right] \right\}+y^2
}
.
\label{eq:26}
\end{eqnarray}
In a similar way, for the TE polarization of the electromagnetic field one finds
\begin{eqnarray}
&&
r^{D(1)}_{\te} (i \zeta_l,y) r^{D(2)}_{\te}(i \zeta_l,y)
=
r^{p(1)}_{\te} (i \zeta_l,y) r^{p(2)}_{\te}(i \zeta_l,y)
\nonumber\\
&&~~~-
\frac{\widetilde{\gamma}^{(1)}(T)}{\zeta_l}
R^{(1)}_{\te} (i \zeta_l,y)
-
\frac{\widetilde{\gamma}^{(2)}(T)}{\zeta_l}
R^{(2)}_{\te} (i \zeta_l,y),
\label{eq:27}
\end{eqnarray}
where
\begin{align}
\begin{split}
&R^{(n)}_{\te} (i\zeta_l,y)
= -
\frac{\beta_{n}\mu_0^{(n)}y }
{
    \sqrt{
        \beta_n^2 y^2 + \beta_n^2 \zeta_l^2
        (\mu_0^{(n)}-1)+\mu_0^{(n)}
    }
}
\\ & \times
\frac{r_{\te}^{p(1)} (i \zeta_l,y)
r_{\te}^{p(2)} (i \zeta_l,y)}
{
\beta_n^2
\left[
(\mu_0^{(n)}-1)
y^2
-
\zeta_l^2
(\mu_0^{(n)}-1)
\right]-\mu_0^{(n)}
}.
\label{eq:28}
\end{split}
\end{align}

The logarithms containing the products~\eqref{eq:25} and~\eqref{eq:27} can
also be expanded in powers of the same small parameters
\begin{align}
\begin{split}
&\ln\left[1-r_{\alpha}^{D(1)} (i\zeta_l,y)r_{\alpha}^{D(2)} (i\zeta_l,y) e^{-y}\right]
\\ &=
\ln\left[1-r_{\alpha}^{p(1)} (i\zeta_l,y)r_{\alpha}^{p(2)} (i\zeta_l,y) e^{-y}\right]
\\ &-
\frac{\widetilde{\gamma}^{(1)}(T)}{\zeta_l}
\frac{R^{(1)}_{\alpha} (i \zeta_l,y)e^{-y}}{1-r_{\alpha}^{p(1)} (i\zeta_l,y)r_{\alpha}^{p(2)} (i\zeta_l,y) e^{-y}}
\\ &-
\frac{\widetilde{\gamma}^{(2)}(T)}{\zeta_l}
\frac{R^{(2)}_{\alpha} (i \zeta_l,y)e^{-y}}{1-r_{\alpha}^{p(1)} (i\zeta_l,y)r_{\alpha}^{p(2)} (i\zeta_l,y) e^{-y}}.
\label{eq:29}
\end{split}
\end{align}

As a result, Eq.~\eqref{eq:24} can be written in the form
\begin{equation}
\F_{D}(a,T)=\F_{p}(a,T)+\F_{0}(a,T)+\F_{\gamma}(a,T),
\label{eq:30}
\end{equation}
where $\F_{0}(a,T)$ is the contribution at zero Matsubara frequency given by
\begin{align}
\begin{split}
\F_{0}(a,T)=&\frac{\kb T}{16 \pi a^2}
\int_{0}^{\infty} y \, dy
\left\{
    \ln\left[1-r^{D(1)}_{\te}(0,y)r^{D(2)}_{\te}(0,y)e^{-y}\right]
    \right. \\ - & \left.
    \ln\left[1-r^{p(1)}_{\te}(0,y)r^{p(2)}_{\te}(0,y)e^{-y}\right]
\right\}
\label{eq:31}
\end{split}
\end{align}
and $\F_{\gamma}$ is the contribution of all nonzero Matsubara frequencies
\begin{align}
\begin{split}
\F_{\gamma}(a,T)&=
-\frac{\kb T}{8 \pi a^2}
\sum_{l=1}^{\infty}
\frac{\widetilde{\gamma}^{(1)}(T)}{\zeta_l}
\\&\times
\int_{\zeta_l}^{\infty} y \, dy
\sum_{\alpha}
\frac{R^{(1)}_{\alpha} (i \zeta_l,y)e^{-y}}{1-r_{\alpha}^{p(1)} (i\zeta_l,y)r_{\alpha}^{p(2)} (i\zeta_l,y) e^{-y}}
\\
&-\frac{\kb T}{8 \pi a^2}
\sum_{l=1}^{\infty}
\frac{\widetilde{\gamma}^{(2)}(T)}{\zeta_l}
\\&\times
\int_{\zeta_l}^{\infty} y \, dy
\sum_{\alpha}
\frac{R^{(2)}_{\alpha} (i \zeta_l,y)e^{-y}}{1-r_{\alpha}^{p(1)} (i\zeta_l,y)r_{\alpha}^{p(2)} (i\zeta_l,y) e^{-y}}.
\label{eq:32}
\end{split}
\end{align}

We consider first the contribution to Eq.~\eqref{eq:30} at zero frequency.
The first integral in Eq.~\eqref{eq:31} contains $r_{\te}^{D(n)}$.
Using Eq.~\eqref{eq:22} it can be calculated explicitly. The second integral in
Eq.~\eqref{eq:31} containing $r_{\te}^{p(n)}$ can be expanded in powers of a small parameter
$\Lambda$ defined in Eq.~\eqref{eq:09}, like this was done in Sec.~\ref{sec:plasma_model}.
Thus Eq.~\eqref{eq:31} can be written as
\begin{align}
\begin{split}
\F_{0}(a,T)=&\frac{\kb T \zeta_{R}(3)}{16 \pi a^2}
\left[
1-\frac{\Li{3}{r_{\mu}^{(1)}r_{\mu}^{(2)}}}{\zeta_{R}(3)}
-4 \Lambda
+12 \Lambda^2
\right].
\label{eq:33}
\end{split}
\end{align}

Then we consider the contribution $\F_{\gamma}$ on the right-hand side of Eq.~\eqref{eq:30}
defined in Eq.~\eqref{eq:32}. Expanding it in powers of small parameters $\beta_n$,
one obtains
\begin{align}
\begin{split}
\F_{\gamma}=&\frac{\kb T}{8 \pi a^2} \sum_{l=1}^{\infty}
\left\{
    \frac{\widetilde{\gamma}^{(1)}(T)}{\zeta_l}
    \int_{\zeta_l}^{\infty} y \, dy
    \left[
        \frac{\beta_1 \zeta_l^2 \sqrt{\mu_0^{(1)}}}{y \left(e^{y}-1\right)}
        +
        \frac{\beta_1 y \sqrt{\mu_0^{(1)}}}{e^{y}-1}
    \right]
    \right.
    \\
    +&
    \left.
    \frac{\widetilde{\gamma}^{(2)}(T)}{\zeta_l}
    \int_{\zeta_l}^{\infty} y \, dy
    \left[
    \frac{\beta_2 \zeta_l^2 \sqrt{\mu_0^{(2)}}}{y \left(e^{y}-1\right)}
    +
    \frac{\beta_2 y \sqrt{\mu_0^{(2)}}}{e^{y}-1}
    \right]
\right\}.
\label{eq:34}
\end{split}
\end{align}
Using Eq.~\eqref{eq:08} and the definition of parameters $\beta_n$, Eq.~\eqref{eq:34} can be presented in the form
\begin{align}
\begin{split}
\F_{\gamma}(a,T)
& =
\frac{\kb T}
{8 \pi a^2}
\left[
\sqrt{\mu^{(1)}_0}
\frac{\gamma^{(1)}(T)}{\omega_p^{(1)}}
+
\sqrt{\mu^{(2)}_0}
\frac{\gamma^{(2)}(T)}{\omega_p^{(2)}}
\right]
\\ & \times
\sum_{l=1}^{\infty}
\int_{\zeta_l}^{\infty} \, d y
\left[
\frac{\zeta_l}{e^y-1}
+
\frac{y^2}{\zeta_l \left(e^y-1\right)}
\right].
\label{eq:35}
\end{split}
\end{align}
Performing summations and integrations in Eq.~\eqref{eq:35}, one arrives at
\begin{align}
\begin{split}
\F_{\gamma}(a,T)
& =
\frac{\kb T_{\text{eff}} \zeta_{R}(3)}
{8 \pi^2 a^2}
\left[
\sqrt{\mu^{(1)}_0}
\frac{\gamma^{(1)}(T)}{\omega_p^{(1)}}
+
\sqrt{\mu^{(2)}_0}
\frac{\gamma^{(2)}(T)}{\omega_p^{(2)}}
\right]
\\ & \times
\left[
    -\ln{\tau}
    +2
    +\frac{\pi^2}{4 \zeta_{R}(3)}
    \tau
\right].
\label{eq:36}
\end{split}
\end{align}
Now we substitute Eqs.~\eqref{eq:33} and~\eqref{eq:36} in Eq.~\eqref{eq:30} and obtain
\begin{eqnarray}
&&
\F_{D}(a,T)
=
\F_{p}(a,T)
+
\frac{\kb T \zeta_{R}(3)}
{16 \pi a^2}
\nonumber\\&&~~\times
\left[
    1
    -
    \frac{\Li{3}{r_{\mu}^{(1)}r_{\mu}^{(2)}}}{\zeta_{R}(3)}
    -4\Lambda
    +12\Lambda^2
\right]
\nonumber\\
&&~~~+
\frac{\kb T_{\text{eff}} \zeta_{R}(3)}{8 \pi^2 a^2}
\left[
    \sqrt{\mu_0^{(1)}}
    \frac{\gamma^{(1)}(T)}{\omega_p^{(1)}}
    +
    \sqrt{\mu_0^{(2)}}
    \frac{\gamma^{(2)}(T)}{\omega_p^{(2)}}
\right]
\nonumber\\
&&~~~\times
\left[
    -\ln{\tau}
    +2
    +\frac{\pi^2}{4\zeta_R(3)}
    \tau
\right].
\label{eq:37}
\end{eqnarray}

It is easily seen that the contribution of $\F_{\gamma}$ to the Casimir entropy goes to
zero with vanishing temperature. For perfect crystal lattices at temperatures
below liquid helium temperature it holds $\gamma^{(n)}(T)=\gamma_0^{(n)}T^2$~\cite{c26,c38}. Then from
Eq.~\eqref{eq:35} we find that
\begin{align}
\begin{split}
\frac{\partial \F_{\gamma} (a,T)}{\partial T}
=&
\frac{\kb T_{\text{eff}} \zeta_{R}(3)}{4 \pi^2 a^2}
\left[
    \sqrt{\mu_0^{(1)}} \frac{\gamma_0^{(1)}}{\omega_p^{(1)}}
    +
    \sqrt{\mu_0^{(2)}} \frac{\gamma_0^{(2)}}{\omega_p^{(2)}}
\right]
\\\times &
\left\{
    T
    \left[
        -\ln{\tau}
        +2
        +\frac{\pi^2}{4 \zeta_{R}(3)}\tau
    \right]
    \right. \\+ & \left.
    T^2
    \left[
        -\frac{1}{\tau}
        +
        \frac{\pi^2}{4\zeta_{R}(3)}
    \right]
    \frac{\pi}{T_{\text{eff}}}
\right\}
\label{eq:38}
\end{split}
\end{align}
and
\begin{equation}
\lim_{T \to 0} \frac{\partial \F_{\gamma} (a,T)}{\partial T} = 0.
\label{eq:39}
\end{equation}

As a result, for the Casimir entropy calculated using the Drude model from
Eq.~\eqref{eq:37} one obtains.
\begin{eqnarray}
&&
S_{D} (a,T) = S_p(a,T)
\label{eq:40}\\
&&~~- \frac{\kb \zeta_{R}(3)}{16 \pi  a^2}
\left[
1
-
\frac{\Li{3}{r_{\mu}^{(1)}r_{\mu}^{(2)}}}   {\zeta_R(3)}
-
4\Lambda
+
12\Lambda^2
\right]
 -
\frac{\partial \F_{\gamma}(a,T)}{\partial T},
\nonumber
\end{eqnarray}
where $S_p$ is given in Eq.~\eqref{eq:19} and $\partial\F_{\gamma}/\partial T$ in Eq.~\eqref{eq:39}.
From Eq.~\eqref{eq:40} it is seen that
\begin{eqnarray}
&&
S_{D} (a,0) =
\lim_{T \to 0} S_{D} (a,T)
\label{eq:41}\\
&&~~=
-
\frac{\kb \zeta_{R}(3)}{16 \pi  a^2}
\left[
    1
    -
    \frac{\Li{3}{r_{\mu}^{(1)}r_{\mu}^{(2)}}}   {\zeta_R(3)}
    -
    4\Lambda
    +
    12\Lambda^2
\right].
\nonumber
\end{eqnarray}

Thus, the Casimir entropy at zero temperature calculated using the Drude model
is not equal to zero and depends on the parameters of our system, such as the
volume (through the separation distance $a$) and the properties of the plates
(through the magnetic permeabilities $\mu_0^{(n)}$ and the plasma frequencies
$\omega_p^{(n)}$). Taking into account that the Casimir entropy is the single
separation-dependent contribution to the total entropy of the closed system, one
arrives to the conclusion that in this case the Nernst heat theorem is
violated~\cite{c39a, c39b}.

It is interesting to analyze the result~\eqref{eq:41} in more detail. In the
case of one magnetic metal we have $r_{\mu}^{(2)}=0$ and Eq.~\eqref{eq:41}
takes a more simple from
\begin{equation}
S_{D} (a,0)=
-
\frac{\kb \zeta_{R}(3)}{16 \pi  a^2}
\left[
1
-
4\Lambda
+
12\Lambda^2
\right]
<0,
\label{eq:42}
\end{equation}
where from Eq.~\eqref{eq:09}
\begin{equation}
\Lambda = \frac{\sqrt{\mu_0^{(1)}}\lambda_p^{{(1)}}+\lambda_p^{(2)}}{4 \pi a}.
\label{eq:43}
\end{equation}
One can see that in this case the Casimir entropy at zero temperature is always
negative. For two nonmagnetic plates $\mu_0^{(1)}=1$ and we reobtain the known result
for two nonmagnetic metals~\cite{c02,c04,c24,c25,c26}.

The most interesting is the case of two magnetic metals. Here, the dependence of
$S_{D}(a,0)$ on the magnetic permeabilities of the plates leads to unexpected results.
For the sake of simplicity, we consider similar plates made of magnetic metal with
the magnetic permeability $\mu_0=\mu_0^{(1)}=\mu_0^{(2)}$. For two similar plates we have
\begin{align}
\begin{split}
r_{\mu} =& \frac{\mu_0-1}{\mu_0+1}
= 1 - \frac{2}{\mu_0+1},\\
r_{\mu}^2 =& 1 - \frac{4 \mu_0}{(\mu_0+1)^2} \approx 1-\frac{4}{\mu_0}.
\label{eq:44}
\end{split}
\end{align}
Taking into account that according to Eq.~\eqref{eq:44}
\begin{equation}
r_{\mu}^{2n} \approx 1 - \frac{4 n}{\mu_0},
\label{eq:45}
\end{equation}
we find
\begin{equation}
\Li{3}{r_{\mu}^2} = \sum_{n=1}^{\infty} \frac{r_{\mu}^{2n}}{n^3}
\approx
\zeta_{R} (3)
-
\frac{2\pi^2}{3 \mu_0}.
\label{eq:46}
\end{equation}
Substituting Eq.~\eqref{eq:46} in Eq.~\eqref{eq:41},
it is easily seen that under the condition
\begin{equation}
\frac{\pi^2}{6 \mu_0\zeta_{R}(3)} < \Lambda
\label{eq:47}
\end{equation}
the entropy at zero temperature is positive. Taking into account the definition
of $\Lambda$ in Eq.~\eqref{eq:09}, one arrives from Eq.~\eqref{eq:47} to an equivalent
condition
\begin{align}
a < \frac{3 \lambda_p \mu_0^{3/2}\zeta_{R}(3)}{\pi^3}.
\label{eq:48}
\end{align}
As an example, for Ni the right-hand side of Eq.~\eqref{eq:48} is equal to
approximately $34\,\mu$m. Thus, for Ni plates the Casimir entropy at $T=0$
is positive over wide range of separations from approximately 2\,$\mu$m
(the application condition of our perturbation approach) to $34\,\mu$m.
This is quite different from the previously investigated case of nonmagnetic
metals.

\section{\label{sec:freq_dependence}The role of frequency dependence of magnetic permeability}

In this section we consider frequency-dependent magnetic permeabilities $\mu^{(n)}(i \xi)$.
It is known ~\cite{c40} that permeability of ferromagnetic metals calculated along the
imaginary frequency axis decreases with the increase of $\xi$ and at some
value $\xi_c$ specific for each metal abruptly drops to unity. As was noted
in Ref.~\cite{c41}, at room temperature the inequality $\xi_1 \gg \xi_c$ holds. Because
of this, in all applications of the Lifshitz theory at $T=300$~\si{\K} one can
put $\mu_l^{(n)}$ at all $l \ge 1$ and take the ferromagnetic properties into account
only in the zero-frequency term $l=0$. This approach was used~\cite{c31,c32,c33} in
the comparison of the theoretical predictions with the measurement data.

The rate of decrease of $\mu^{(n)}(i \xi)$ with increasing $\xi$ depends on
the value of electric resistance. The lower is the resistance of
a ferromagnetic material, the lower is the frequency $\xi_c$ at which $\mu^{(n)}(i \xi)$
drops to unity~\cite{c40}. For typical ferromagnetic metals $\xi_c$ is of
order $10^5$ Hz. Thus, not only at room temperature, but even at relatively
low temperature $T>0.001$K one can put $\mu^{(n)}(i \xi)=1$ for all $l \ge 1$. This means
that for real ferromagnetic metals the thermal correction to the Casimir
energy calculated using the plasma model can be presented in the form
\begin{align}
\begin{split}
& \Dt \F(a,T) = \frac{\hbar c}{16 \pi^2 a^3}
\sum_{l=1}^{\infty}
\left[
    B_l^{(0)}(t)+B_l^{(1)}(t)\Lambda_1+B_l^{(2)}(t)\Lambda^2_1
\right]
\\ & +
\frac{\kb T}{16 \pi a^2}
\left\{
\int_{0}^{\infty} y \,dy \ln \left[
        1
        -
        \frac{\mu_0^{(1)}y - \sqrt{\mu_0^{(1)}(\widetilde{\omega}_p^{(1)})^2+y^2}}
        {\mu_0^{(1)}y + \sqrt{\mu_0^{(1)}(\widetilde{\omega}_p^{(1)})^2+y^2}}
        \right. \right. \\ & \times \left.
        \frac{\mu_0^{(2)}y - \sqrt{\mu_0^{(2)}(\widetilde{\omega}_p^{(2)})^2+y^2}}
        {\mu_0^{(2)}y + \sqrt{\mu_0^{(2)}(\widetilde{\omega}_p^{(2)})^2+y^2}}
        e^{-y}
    \right]
    \\ & - \left.
    \int_{0}^{\infty} y \,dy \ln
    \left[
        1
        -
        \frac{y - \sqrt{(\widetilde{\omega}_p^{(1)})^2+y^2}}
        {y + \sqrt{(\widetilde{\omega}_p^{(1)})^2+y^2}}
        \,
        \frac{y - \sqrt{(\widetilde{\omega}_p^{(2)})^2+y^2}}
        {y + \sqrt{(\widetilde{\omega}_p^{(2)})^2+y^2}}
        e^{-y}
    \right]
\right\},
\label{eq:49}
\end{split}
\end{align}
where
\begin{equation}
\Lambda_1 = \frac{\lambda_p^{(1)}+\lambda_p^{(2)}}{4 \pi a} \ll 1.
\label{eq:50}
\end{equation}
The sum on the right-hand side of Eq.~\eqref{eq:49} is the thermal
correction~\eqref{eq:15}, where the magnetic properties are omitted
[this is seen from the replacement of $\Lambda$ defined in Eq.~\eqref{eq:09} with $\Lambda_1$
defined in Eq.~\eqref{eq:50}]. The following two integrals add the contribution of
the zero-frequency term with included magnetic properties to the thermal
correction (the first one) and subtract the contribution of the same term with
omitted magnetic properties. [Note that the replacement of $\mu^{(n)}(i \xi)$ with unity in
the frequency region [0,$\xi_c$] leads to only a negligibly small influence
on the value of $E(a)$.] In Eq.~\eqref{eq:49} we have also taken into account
that $r_{\tm}^{(n)}(0,y)=1$ for both magnetic and nonmagnetic materials. As a result, only
a difference in the values of $r_{\te}^{(n)}(0,y)$ for $\mu=\mu_0^{(n)}$ and $\mu=1$ contributes to $\Dt \F(a,T)$.

We expand the first and second integrals in the right-hand side of Eq.~\eqref{eq:49}
in powers of small parameters $\Lambda$ and $\Lambda_1$, respectively, and perform
integrations with respect to $y$. The result is
\begin{align}
\begin{split}
\Dt \F(a,T) = & \frac{\hbar c}{16 \pi^2 a^3} \sum_{l=1}^{\infty}
\left[
B_l^{(0)}(t)+B_l^{(1)}(t)\Lambda_1+B_l^{(2)}(t)\Lambda^2_1
\right]
\\+&
\frac{\pi \kb T}{48 a^2} \left(\Lambda-\Lambda_1\right)
\left[
    1-2\left(\Lambda+\Lambda_1\right)
\right].
\label{eq:51}
\end{split}
\end{align}
For the case of similar plates we have
\begin{align}
\Lambda \pm \Lambda_1 = \left( \sqrt{\mu_0} \pm 1\right) \frac{\lambda_p}{2\pi a}.
\label{eq:52}
\end{align}
Note that alternatively we can do not expand the integrals on the right-hand side of
Eq.~\eqref{eq:49} in powers of small parameters, but calculate them numerically.
In this case the application region of Eq.~\eqref{eq:49} is determined by the
condition $\Lambda_1 \ll 1$. For example for Ni this leads to $a>200$~\si{\nm} [i.e. Eq.~\eqref{eq:49}
is applicable starting from an order of magnitude smaller separations than Eq.~\eqref{eq:51}].

One can use Eqs.~\eqref{eq:49} and~\eqref{eq:51} to calculate the thermal correction to the Casimir energy at $T>0.001$K. From Eq.~\eqref{eq:49} and Eq.~\eqref{eq:14} it is easy also to obtain the respective expression for the thermal correction to Casimir pressure between two parallel plates made of ferromagnetic metals
\begin{eqnarray}
&&
\Dt P(a,T) = - \frac{\partial}{\partial a} \Dt \F(a,t)
\nonumber\\
&&~~ = - \frac{\hbar c}{8 \pi^2 a^4}
\sum_{l=1}^{\infty}
\biggl\{
    \frac{1}{\left(l t\right)^4}
    -
    \frac{\pi^3}{lt}
    \frac{\coth{\pi lt}}
    {\sinh[2]{\pi lt}}
\nonumber    \\
   &&~~~+
    \Lambda_1 \frac{\pi^3}{lt \sinh[2]{\pi l t}}
    \left[
        \frac{1}{\left(\pi l t\right)^2}
        \sinh{\pi l t} \cosh{\pi l t}
        \right.
\nonumber    \\
&&~~~
+ 4 \coth{\pi l t}
        + \left.  2 \pi l t - 6 \pi l t \coth[2]{\pi l t} + \frac{1}{\pi l t}
    \right]
\nonumber    \\
&&~~~
+
    3 \Lambda_1^2 \frac{\pi^3}{lt \sinh[2]{\pi l t}}
    \left[
        -4 \pi l t
        +
        5 \left(\pi l t \right)^2
        \coth{\pi l t}
    \right]
\nonumber    \\
&&~~~+
    12 \pi lt \coth[2]{\pi l t} - 8 \left(\pi l t \right)^2 \coth[3]{\pi lt }
     -  4 \coth{\pi l t}
\biggr\}
\nonumber\\
&&~~~+
\frac{\pi \kb T}{16 a^3}
\left(\Lambda - \Lambda_1\right)
\left[1-\frac{4}{3} \left(\Lambda+\Lambda_1\right)\right].
\label{eq:53}
\end{eqnarray}

We remind that it is not possible to consider the limiting case $T \to 0$ in
Eqs.~\eqref{eq:49},~\eqref{eq:51} and~\eqref{eq:53} because these equations are
obtained under a condition that the temperature is larger than some fixed (small)
value. In order to investigate the role of the frequency dependence
of $\mu(i \xi)$ at $T \to 0$ it is convenient to use the Abel-Plana formula similar
to Ref.~\cite{c42} (see also Ref.~\cite{c02})

\begin{equation}
\sum_{l=0}^{\infty}\phantom{}^{'}
\Phi(l) = \int_{0}^{\infty} \Phi(t)\, dt
+
i
\int_{0}^{\infty} \, dt
\frac{\Phi(it)-\Phi(-it)}{e^{2\pi t}-1}.
\label{eq:54}
\end{equation}
Now we choose
\begin{equation}
\Phi(\zeta) = \int_{0}^{\infty} y \,dy F(\zeta,y),
\label{eq:55}
\end{equation}
where $F(\zeta,y)$ is defined in Eq.~\eqref{eq:05}, and obtain the thermal
correction to the Casimir energy in the form~\cite{c02,c42}
\begin{equation}
\Dt \F(a,t) = \frac{i \hbar c \tau}{32 \pi^2 a^3}
\int_{0}^{\infty} \, dt
\frac{\Phi(i \tau t)-\Phi(-i \tau t)}
{e^{2\pi t} - 1}.
\label{eq:56}
\end{equation}

The dependence of the magnetic permeability $\mu(i\xi)$ on the frequency
is described by the Debye formula~\cite{c40}
\begin{equation}
\mu(i \zeta) = 1 + \frac{\mu_0 - 1}{1+ \frac{\xi}{\omega_m}}
=
1+\frac{\mu_0-1}{1+\ae_m\zeta},
\label{eq:57}
\end{equation}
where $\ae_m=\omega_c/\omega_m$ and $\omega_m$ is some characteristic frequency which is different
for different materials. For simplicity we consider the case of two similar
magnetic metals and restrict ourselves by the first order terms in $\Lambda$.
Then from Eq.~\eqref{eq:55} we obtain
\begin{align}
\begin{split}
\Phi(\zeta) &= 2 \int_{\zeta}^{\infty} y \, dy \left[\ln\left(1-e^{-y}\right)
+
\frac{\zeta^2+y^2}{y\left(e^y-1\right)}
\Lambda
\right]
\\ &-
2\frac{\left(\mu_0-1\right) \ae_m}{\sqrt{\mu_0}}
\zeta \Lambda
\int_{\zeta}^{\infty}
\frac{\zeta^2+y^2}{e^y-1}
\, dy.
\label{eq:58}
\end{split}
\end{align}
The contribution of the first integral on the right-hand side of Eq.~\eqref{eq:58}
to $\Dt\F(a,T)$ was calculated in Ref.~\cite{c42}. Because of this,
now we consider only the second one
\begin{equation}
\Phi_2(\zeta) \equiv -2 \frac{\left(\mu_0-1\right)\ae_m}{\sqrt{\mu_0}}
\zeta \Lambda \int_{\zeta}^{\infty} \frac{\zeta^2+y^2}{e^y-1}\, dy.
\label{eq:59}
\end{equation}
In the lowest order to the small parameter $\zeta=t\tau$, we have
\begin{equation}
\int_{\zeta}^{\infty} \frac{\zeta^2+y^2}{e^y-1} \, d y =
\zeta_{R}(3)+O(\zeta).
\label{eq:60}
\end{equation}
Then, in the lowest order in $\tau$ one obtains
\begin{equation}
\Phi_1(i \tau t) - \Phi_1(-i \tau t)
=
- 4 i
\frac{\left(\mu_0-1\right)\ae_m \Lambda \zeta_{R}(3)}{\sqrt{\mu_0}} \tau t.
\label{eq:61}
\end{equation}
Substituting this in Eq.~\eqref{eq:56} together with the result of Ref.~\cite{c42}
for the first integral on the right-hand side of Eq.~\eqref{eq:58}, we arrive at
\begin{align}
\begin{split}
\Dt \F(a,T) &= - \frac{\hbar c}{ 8 \pi a^3}
\left\{
    \frac{\zeta_R(3)}{2 t^3}
    -
    \frac{\pi^3}{90 t^4}
    +
    \Lambda
    \left[
        \frac{\zeta_{R}(3)}{t^3}
        -
        \frac{2 \pi^3}{45 t^4}
    \right]
    \right.
    \\&-
    \left.
    \frac{\zeta_{R}(3)}{6 \pi}
    \frac{\left(\mu_0-1\right)\ae_m \Lambda}{\sqrt{\mu_0} t^2}
\right\}.
\label{eq:62}
\end{split}
\end{align}
This result is in agreement with Eq.~\eqref{eq:17}, but contains an additional
term due to the frequency dependence of $\mu$.

Thus, for magnetic metals, an account of the frequency dependence of $\mu$ at
small $T$ gives rise to the second order in $T$ correction in the free energy.
As a result, the entropy~\eqref{eq:19} acquires a correction
\begin{equation}
\Delta S(a,T) = - \frac{\kb \zeta_{R} (3) \left(\mu_0-1\right) \ae_m \Lambda}
{24 \pi^3 a^2 \sqrt{\mu_0}}
\tau.
\label{eq:63}
\end{equation}
As is seen in Eq.~\eqref{eq:63}, this correction goes to zero when $T$ goes to zero, i.e.,
for the plasma model the Nernst heat theorem is preserved even with account of
the frequency dependence of $\mu$. This is in analogy to the case of magnetodielectrics
investigated in Ref.~\cite{c43}.

For the Drude model, an account of the frequency dependence of $\mu$ does not
change the fact that for metals with perfect crystal lattices the Nernst heat
theorem is violated. This follows from Eq.~\eqref{eq:30}, where the violation of the Nernst
theorem originates from the contribution of the zero frequency terms entering $\F_0(a,T)$.
For $\F_p(a,T)$ on the right hand side of Eq.~\eqref{eq:30} the Nernst theorem is satisfied for the
frequency-dependent $\mu$ and $\F_0(a,T)$ does not depend on the presence of the frequency
dependence. Finally, the term $\F_{\gamma}(a,T)$ acquires a correction of higher order in $T$
to the terms written in Eq.~\eqref{eq:36}. This proves that the frequency dependence of
$\mu$ does not change our conclusions concerning a consistency of the plasma and
Drude model approaches with thermodynamics for the case of magnetic metals.

\section{\label{sec:conc}Conclusions and discussion}

In the foregoing, we have investigated analytic behavior of the Casimir free
energy and entropy between two parallel plates made of dissimilar ferromagnetic
metals. In so doing, the dielectric properties of metals were described
either by the nondissipative plasma model or by the Drude model taking into account
the dissipation of free charge carriers at arbitrarily low temperature.

Using the perturbation expansions of the Casimir free energy in small parameters
it was shown that the Lifshitz theory combined with the plasma model satisfies
the Nernst heat theorem, i.e., the Casimir entropy goes to zero when
the temperature vanishes. Quite differently, for ferromagnetic metals with
perfect crystal lattices described by the Drude model it was shown that in the limit
of zero temperature the Casimir entropy goes to a nonzero limit depending
on the parameters of the system, i.e., the Nernst heat theorem is violated.
Both constant and frequency dependent magnetic permeabilities are considered.

As was noted in Sec.~\ref{sec:intro}, violation of the Nernst heat theorem for
the Casimir entropy between nonmagnetic metals with perfect crystal lattices
is a known effect~\cite{c02,c04,c24,c25,c26}. The distinctive feature of magnetic
metals found in this paper is that the Casimir entropy at zero temperature
depends on the static magnetic permeabilities of plate metals and can be
positive (recall that for nonmagnetic metals with perfect crystal lattices
described by the Drude model the Casimir entropy at $T=0$ is always negative).
This establishes a link between ferromagnetic metals described by the Drude
model and dielectrics with taken into account dc conductivity of plate materials.
In the latter case it is known~\cite{c02,c04,c44,c45} that the Lifshitz theory violates
the Nernst heat theorem and the Casimir entropy at $T=0$ is positive.
For magnetic dielectrics this positive quantity remains independent on the magnetic
properties~\cite{c43}. This makes unique the case of ferromagnetic metals
considered here. Note also that we have considered the configuration of two
parallel plates, but a violation of the Nernst heat theorem for the Casimir entropy
calculated using the Drude model holds also for other configurations.
Thus, for a sphere above a plate made of nonmagnetic metals this was demonstrated in Ref.~\cite{c46}.

It is pertinent to briefly discuss the physical meaning of the Casimir entropy and
its sign. For this purpose we remind that the Casimir free energy per unit area of plates
(\ref{eq:01}) is derived by subtraction of the free energy for infinitely separated
(uncoupled) plates from the nonrenormalized free energy of plates separated by a distance
$a$ \cite{c02}. Then the Casimir entropy calculated by Eq.~(\ref{eq:18}) also represents
respective difference and has the meaning of an entropy of the fluctuating field.
It characterizes an interaction between the plates, but considers only a minor fraction
of the entropy of a closed system which includes also much larger entropies of the plates.
As a result, at not too low temperature the total entropy is always positive irrespective of
whether the Casimir entropy is positive or negative. An important point, however, is that
only the Casimir entropy depends on the separation distance, whereas the entropies of
the plates do not depend on separation. Because of this, the total entropy at $T=0$ is
separation-dependent, i.e., the Nernst heat theorem is violated.
In fact the Casimir entropy per unit area of two parallel plates is not an immediately
measured quantity. It can be experimentally found indirectly by means of numerical
differentiation from the force between a sphere and a plate measured as a function of
temperature (in the Derjaguin approximation \cite{c02} the latter quantity is proportional
to the Casimir free energy per unit area of two parallel plates).

As was mentioned in Sec.~\ref{sec:intro}, the most precise experiments using
nonmagnetic metals~\cite{c19,c20,c21,c22,c23} and all the experiments using
magnetic metals~\cite{c31,c32,c33,c35} are in agreement with the plasma model
approach and exclude the Drude model approach to calculation of the Casimir force. In a similar way, the measurement
data of most precise experiments using dielectric test bodies~\cite{c09,c15,c16,c47,c48,c49}
are in agreement with theoretical predictions of the Lifshitz theory only
if the dc conductivity of boundary materials is omitted in calculations. In our opinion,
it cannot be accidental that in so many experiments on measuring the Casimir
interaction the data are in agreement with thermodynamically consistent theoretical
approach and exclude the approaches where the Nernst heat theorem is violated.
Thus, the problem of a proper account of the relaxation properties of free charge carriers
in metals and the dc conductivity in dielectrics when calculating the Casimir
interaction invites further investigation.

\begin{acknowledgments}
The authors are grateful to V.~M. Mostepanenko for useful discussions.
\end{acknowledgments}


\end{document}